\documentclass[11pt]{article}

\usepackage[margin=1in]{geometry}

\usepackage{natbib}
\bibpunct[, ]{(}{)}{,}{a}{}{,}%

\usepackage{booktabs}
\usepackage{amsmath}
\usepackage{amsfonts}
\usepackage{amsthm}               
  {
      \newtheorem{assumption}{Assumption}
      \newtheorem{lemma}{Lemma}
      \newtheorem{proposition}{Proposition}
      
      \newtheorem{theorem}{Theorem}
      \newtheorem{example}{Example}
  }
\usepackage[lofdepth,lotdepth]{subfig}
\usepackage[colorlinks=true, allcolors=blue]{hyperref}
\usepackage[titletoc,title]{appendix}
\usepackage{graphicx}
\usepackage{algorithm}
\usepackage[noend]{algpseudocode}
\usepackage{bbm}
\usepackage{authblk}

\newcommand{\st}{\text{  s.t. }}
\newcommand{\andd}{\text{ and }}

\begin{document}

\title{The Second-Price Knapsack Problem: Near-Optimal Real Time Bidding in Internet Advertisement}

\renewcommand*{\Affilfont}{\itshape\small}
\author[1]{Jonathan Amar\thanks{amarj@mit.edu}}	
\author[1]{Nicholas Renegar\thanks{nrenegar@mit.edu}}	

\affil[1]{Massachusetts Institute of Technology - Operations Research Center}	
\date{}
\maketitle

\begin{abstract}
  In many online advertisement (ad) exchanges, ad slots are each sold via a separate second-price auction. This paper considers the bidder's problem of maximizing the value of ads they purchase in these auctions, subject to budget constraints. This 'second-price knapsack' problem presents challenges when devising a bidding strategy because of the uncertain resource consumption: bidders win if they bid the highest amount, but pay the second-highest bid, unknown a priori. This is in contrast to the traditional online knapsack problem, where posted prices are revealed when ads arrive, and for which there exists a rich literature of primal and dual algorithms. 
  
  The main results of this paper establish general methods for adapting these primal and dual online knapsack selection algorithms to the second-price knapsack problem, where the prices are revealed only after bidding. In particular, a methodology is provided for converting deterministic and randomized knapsack selection algorithms into second-price knapsack bidding strategies, that purchase ads through an equivalent set of criteria and thereby achieve the same competitive guarantees. This shows a connection between the traditional knapsack selection algorithm and second-price auction bidding algorithms, that has not previously been leveraged.
  
  Empirical analysis on real ad exchange data verifies the usefulness of this method, and gives examples where it can outperform state-of-the-art techniques.
\end{abstract}

\newpage
\section{Introduction}
Online advertising is one of the fastest growing sectors in the Information Technology (IT) industry. Total digital ad spending in the U.S. increased by 16\% year-over-year in 2016 to \$83 billion, and the global digital advertising market is projected to reach a total of \$330 billion by 2020.\footnote{Digital advertising spending worldwide from 2015 to 2020 (in billion U.S. dollars) \textit{https://www.statista.com/statistics/237974/online-advertising-spending-worldwide/}} The increasing volume of web traffic led to the birth of ad exchanges, resulting in a larger market where advertisers have a stronger chance of locating a preferential ad-context and publishers generate more revenue by being matched with these advertisers. 

\begin{centering}
	\begin{figure}[h]	
		\caption{RTB Flow Chart\label{fig:mec}}
		\centering	\includegraphics[width=440pt]{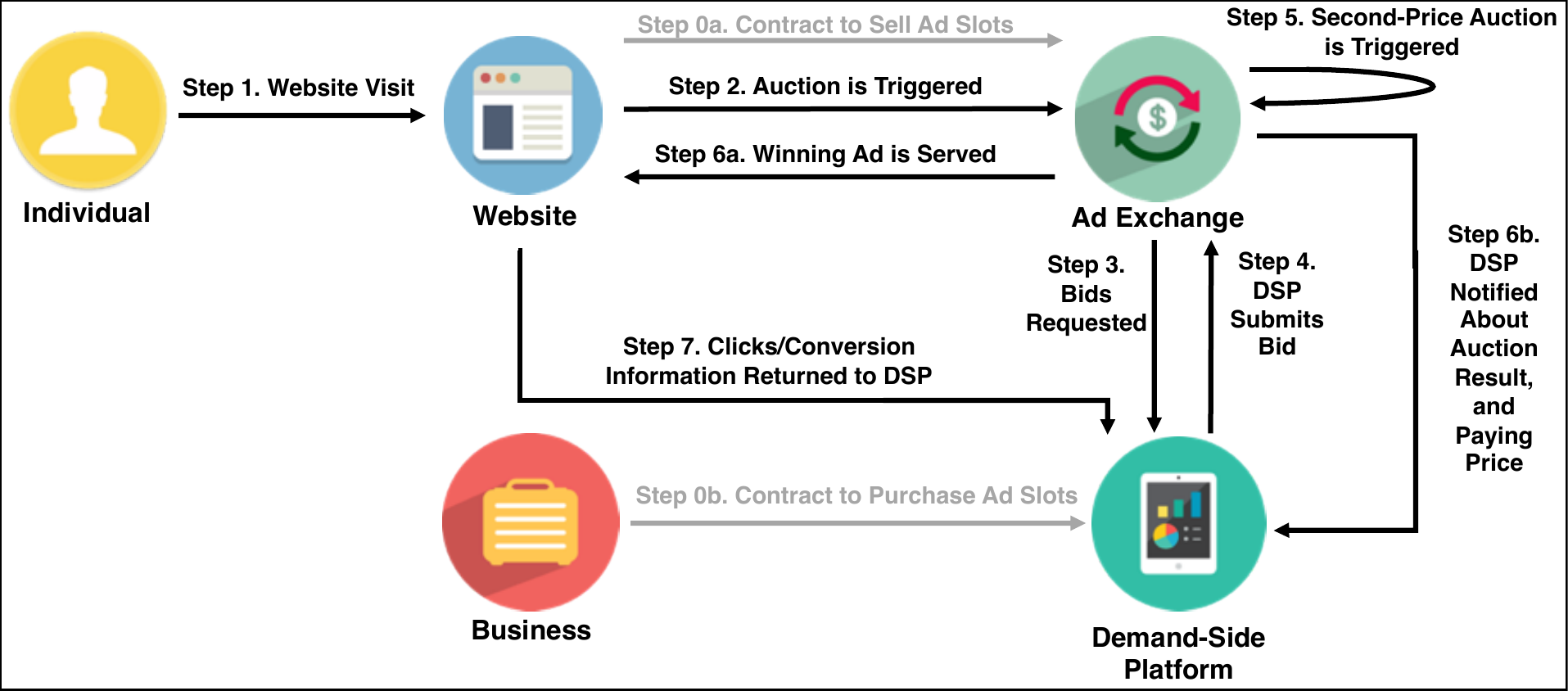}	
	\end{figure} 
\end{centering}

These ad exchanges use a variety of auction mechanisms. One popular auction framework is real-time bidding (RTB). Under the RTB framework, an auction for each ad slot is triggered when a user visits a web-page, with the auction containing contextual parameters; the ad exchange then solicits bid requests from several Demand Side Platforms (DSPs), who each can return a bid for an advertiser it represents; finally the winning ad reaches the publisher. The paying price in RTB is set according to a second-price auction mechanism, i.e. the winner is the bidder with the highest bid, provided that it is above the floor price set by the exchange, and the winner pays a maximum of the floor price and the second highest bid. Given that the auction must be completed before the web-page loads, this imposes latency requirements for the DSP. Notably the bid price calculated and submitted by a DSP must be done within about 10 milliseconds, requiring simple bidding strategies. Furthermore, DSPs may receive a large number of bid requests from exchanges per second, while billions of people explore the web around the globe. Hence, a DSP's job can be quite intensive, and there is a limit to the complexity/number of updates that can be made to the bidding strategy in an online setting. We represent the process of online advertising, bidding and ad allocation in Figure \ref{fig:mec}.

This paper considers the budget constrained bidding problem from the perspective of a bidder (DSP) who represents a single advertiser, a common scenario for large companies. The bidder wants to maximize the total value of ads (indexed by $i \in {I}$) they purchase, where the value of each ad $v_i$ is known to the bidder using contextual parameters (user and ad slot information), subject to a budget constraint $B$. The bidder must create a bidding policy where they bid $\pi_i$ based on the information available to them at the time of the auction: the value of the ad $v_i$, the results of the previous auctions, and their remaining budget. Unlike a traditional knapsack problem, there are two significant difficulties when finding an optimal bidding policy. First, the bidder does not know whether they will win an auction based on the amount that they bid, since the paying price $b_i$ is not determined before completion of the auction. Second, even if they win the auction, they do not know beforehand the amount that they pay: they only know that it will be less than the amount they bid. We generalize this framework to a second-price knapsack problem with uncertain budget consumption in (\ref{eq: kpi-opt}). We refer to this problem as the \textbf{second-price knapsack problem}.

\subsection{Our Contribution}
The main thesis of this paper is that myriad primal and dual algorithms for the traditional online knapsack problem, where participants choose whether to select items with known prices as they arrive, can be adapted to the budget-constrained second-price knapsack problem, where participants must bid on items with unknown prices. In particular, we provide a methodology for converting deterministic (dual) and randomized (primal) online knapsack selection algorithms into second-price knapsack bidding strategies offering the same competitive guarantees. This establishes a strong connection that has not been mentioned in the literature, and yields competitive algorithms for advertising companies (among others) that are readily implementable. 

The paper starts by building an intuition for why this connection might be expected. To do so, we study the retrospective offline setting, in which the set of ads and paying prices $b_i$ are known. Considering this as a selection problem rather than a bidding problem, it reduces to a traditional knapsack problem. From the offline greedy knapsack algorithm, a threshold policy on the value-to-price ratio $\frac{v_i}{b_i}$ is optimal for the linear relaxation of this selection problem which is a known result. That is, we select all ads where $\frac{v_i}{b_i}>\lambda^*$, and some fraction where $\frac{v_i}{b_i}=\lambda^*$, where $\lambda^* \in \mathbb{R}$ is an optimal solution to the dual problem. In our specific context, where $B \gg b_i$, the greedy approach yields a near-optimal solution to the integer selection problem as well. The question is whether this can be turned into an implementable bidding strategy in the second-price knapsack setting, where the information $b_i$ is not available at the time of bidding. Because of the structure of the second-price auction mechanism, we can recover exactly this near-optimal selection with a linear form of bidding only based on the value of the ad.

We then consider the online second-price knapsack problem with budget constraints and known time horizon. We show that under common assumptions (stable arrivals of bid requests, and stable reserve prices and competitors' bidding strategies) we recover the random-permutation assumption that many online-knapsack algorithms rely on. We then establish our main contribution: using the structure of a second-price auction, we show how one can adapt very general classes of primal and dual online knapsack selection algorithms, and apply them to this second-price knapsack problem with uncertain budget consumption. Concretely, we provide a methodology for converting deterministic and randomized selection algorithms into bidding strategies for the second-price knapsack problem. These adaptations will, in effect, select ads according to the same mechanisms, and therefore achieve the same competitive ratios.

We provide examples of these second-price knapsack algorithms, which give us a competitive ratio of $1-\epsilon$, where $\epsilon$ is small in practice and sublinear with respect to the number of ads that arrive. This yields a theoretical result on an important variation of the standard online knapsack problem which arises in the context of Internet advertisement. 

Finally we use the iPinYou dataset to give numerical support to our work, and contrast with current state-of-the-art adaptive pacing algorithms. iPinYou is currently the largest DSP in China. The dataset contains logs of features, bids, assignments, feedback for all impressions over a season. We process the contextual information for every impression, score the market price and customer feedback (click). From reviewing the data, the different features clearly affect the value of an impression which also changed based on the advertiser. We fit a model for each advertiser to estimate the value of an ad based on the contextual features. We evaluate the performance of our example bidding strategy in the online setting and compare it to both adaptive pacing, and the offline optimum value for different advertisers and budgets. We find that for bidders with the budget to purchase only $\frac{1}{16}$ of the ads they are interested in, the example algorithm that is provided performs near-optimally and can outperform adaptive pacing, which is a popular approach both in the literature and in practice, by an average of 25.9\%. These results demonstrate the practical implications of the work.

While this paper focuses on the second-price knapsack mechanism, the theorems also apply to all deterministic, single slot, dominant strategy incentive compatible (DSIC) mechanisms\footnote{For any deterministic DSIC auction, and given the actions of the other participants, there exists a price $p$ for which the bidder wins whenever they bid above $p$, and if so they pay price $p$. It can easily be seen that the main results hold, with identical proofs.}. This encompasses many of the mechanisms that are used in practice for the online ad industry (with some notable exceptions being randomized mechanisms and variations of first-price auctions).
The paper's results have a significant practical impact for online advertisers, by recognizing connections between online selection algorithms for the traditional knapsack problem and bidding algorithms for the second-price knapsack problem, that have not traditionally been leveraged. 

\subsection{Related Work}	
Knapsack problems and the design and analysis of online algorithms have been widely studied in operations research, while the specific RTB context has been studied more within the computer science community. We do not review the broader ad allocation and planning problems which can be combined to our work on bidding strategies.

Our bidding policies are based on a theoretical analysis of the knapsack problem, and its variant the packing problem. In fact, we cast the optimal bidding policy as the solution to a knapsack problem with budget uncertainty, as stated in (\ref{eq: kpi-opt}). We refer the interested reader to \cite{kellerer2004introduction} for a thorough review of knapsack problems. As for the online knapsack problem, \citet{marchetti1995stochastic} prove that the competitive ratio can be pushed away from near-optimality to $1-\frac{1}{e}$ in the adversarial knapsack problem, even for randomized algorithms. \citet{lueker1998average} designs a value-to-bid threshold function which depends on the ratio budget spent over leftover time to design a binary decision function. They then provide a scheme for approximating the optimal thresholding function from observed realizations. The fractional knapsack problem has been studied in \cite{noga2005online}. In our work we use packing duality theory, through which we are able to derive the desired optimality guarantees.

In order to avoid the adversarial setup of the online problem, recent literature considered variations of the stochastic setting. \citet{kleywegt1998dynamic} analyze the stochastic case where items are drawn from an i.i.d. distribution. Under this framework, extensive research has been done for the bandits with knapsacks problem. Further under the random permutation model, \citet{devanur2009adwords} prove a competitive ratio for their algorithm, which is based on a \textit{linear program} (LP). Their approach solves offline the LP associated to a sample of the items, which is large enough to recover the distribution and thus provide concentration inequalities. These then provide guarantees on the dual variables from the sub-sampled LP to the true LP. Integer optimization tools then bridge the gap to recover the true solution. \citet{feldman2010online, agrawal2014dynamic} use similar ideas, but using a dynamic algorithm solving multiple LPs in online fashion. This approach yields tighter bounds given the sharper estimate of the dual parameters from the data. In their work, they focus on providing a competitive ratio analysis for when the input size is large. 

On the other hand, \citet{buchbinder2009design} provide a wide variety of algorithms for more general online optimization problems. They approach these by using Primal-Dual algorithms which yield arbitrarily good competitive ratios while violating dual constraints by some factor. These compare to thresholding functions by setting dual prices for budget consumptions. \citet{babaioff2008online}
analyze the special case of the secretary problem where weights are in $ \{0,1\} $. \citet{buchbinder2005online, buchbinder2006improved} give an algorithm with a multiplicative competitive ratio of $ O(\log(U/L))$ for the online knapsack problem based on a general online primal-dual framework where $U,L$ are respectively constraining upper and lower bounds on the individual value of ads, the resulting competitive ratio scales with the bounds. These algorithms were applied to second-price auctions by \citep{chakrabarty2007budget,zhou2008algorithm}, although the analysis focuses on worst-case guarantees in the adversarial setting, as opposed to our work which studies expected value guarantees in the random permutation setting. Important work on the knapsack problem has also been done in the setting where values are unknown. \citet{badanidyuru2018bandits} study this setting, and develop primal-dual algorithms with sublinear regret. 

We now relate the knapsack problems to the RTB literature. The design of a bidding strategy requires an algorithm to cast real-time decisions for the bidder based on contextual information. Recent contributions have been made to the bidding strategies by formulating the problem as an online knapsack problem. Notably in the advertisement community, \citet{chakrabarty2007budget} design the problem by assuming bounds on items' values, which are small compared to the total budget, under adversarial arrivals. Similarly to \citet{lueker1998average}, they design an online algorithm based on a threshold function guiding their bidding strategy. Their algorithms depend on some input parameters which directly influences its performance.
We extend their work by proving near-optimality of a threshold based algorithm under some large-scale assumptions. The idea of using a threshold function was first introduced by \cite{williamson1992airline}, however they design an asymptotically optimal strategy only when the arrival distribution is known.

Much of the RTB literature makes very stylized and stringent assumptions to provide tractability. These may be assumptions on the distribution of the arrivals or assuming that the probability of winning an auction is a direct function of the amount bid. Keyword auctions have been studied while others deal with more empirical and data driven questions \citep[see][]{ edelman2007strategic, zhou2007vindictive, chen2011real, lee2012estimating, lee2013real, zhang2014real, yuan2013real, ren2018bidding}. There is also relevant work in the stochastic setting, by \citet{jiang2014bidding}, which uses mean-field approximation to model competitors' behaviors and creates an algorithm that converges to an optimal solution in expectation.

The most similar work to our own is recent work on adaptive pacing for the online second-price knapsack problem under budget constraints, when the valuation distributions for different bidders are unknown and independent \cite{balseiro2017learning}. Their algorithm tries to learn the dual parameter in order to keep the expenditure path stable over time, which is an approach we leverage as well. The adaptive pacing algorithm is shown to offer asymptotic optimality for both the stable and arbitrary settings, and converges to an equilibrium in dynamic strategies. Existence of an \textit{asymptotically optimal} constant competitive ratio, versus the offline oracle, is also shown. Unlike our results in Section \ref{sec:onlineselection}, though, a competitive ratio is not explicitly given the finite setting.

\section{Model and Analysis}
	
We take the bidder position of a DSP representing an advertiser. When we are presented with an auction for a new ad slot, we must decide whether to bid on the ad or not, and if so, how much to bid. We consider the set of all non-anticipatory bidding policies $\Pi$, which satisfy our budget constraint, where for any $\pi \in \Pi$ and ad $i$ we bid some amount $\pi_i \ge 0$. We win the auction when $\pi_i$ is the highest bid (and above the floor/reserve price defined for the auction, although this can be considered as another bidder without loss of generality), and pay the maximum of the other bids. We label this paying price $b_i$, and note that we win the auction whenever $\pi_i \ge b_i$ and always pay $b_i$ when we win. Upon winning the auction, 
we collect value $ v_i $, known prior to the auction. The objective is to design a bidding strategy that maximizes the total collected value under budget constraint $ B $.
	
Our main difficulty in designing a bidding strategy for the online setting is that we do not know the paying price $b_i$ a priori. We first consider the offline traditional knapsack problem in which $b_i$ is known. We then show there exists a simple near-optimal bidding policy $\pi$ for the online second-price auction setting where $b_i$ is unknown which is a linear function only of the value of the ads $v_i$ and some bid parameter $\lambda^*$.

For the online problem, we then provide our main results in Theorems \ref{thm:dualOnline} and \ref{thm:primalOnline}, which establish a methodology for adapting a variety of selection algorithms from the traditional online knapsack problem to bidding algorithms in the second price knapsack problem. Using these theorems, we reconsider the specific problem of training the bid parameter $\lambda^*$ for our second-price knapsack strategy, in an online setting. To illustrate this, we show how work by \cite{agrawal2014dynamic} for an online one-shot-learning algorithm with a competitive ratio of $1-6\epsilon$, can be applied to the online second-price knapsack problem where the paying prices $b_i$ are unknown, achieving the same competitive ratio. This is summarized in Example \ref{cor:Gen-RTB}.

\subsection{Model Framework}
We consider the model with respect to a bidding policy $\pi \in \Pi$ satisfying the budget constraint, where we bid $\pi_i$ for ad $i$. Although the paying price $b_i$ is unknown, we win if $\pi_i \ge b_i$, and pay amount $b_i$. This allows us to cast the second-price knapsack problem with budget uncertainty, hence labeled (\ref{eq: kpi-opt}), for the online setting as:
\begin{align}
\label{eq: kpi-opt} \tag{K-2}
&\max_{\pi \in \Pi} \sum_i  \mathbbm{1}[\pi_i \ge b_i] \times v_i  
\st  \sum_i \mathbbm{1}[\pi_i \ge b_i] \times {b}_i  \leq B \: \text{a.s.} \nonumber
\end{align}

In the offline setting where the paying price $b_i$ is known at the time of bidding, the problem reduces to a traditional knapsack problem. Items arrive with known value $v_i$ and paying price $b_i$. If we want to select this item we just bid some amount $\pi_i$ at least as large as $b_i$ and pay $b_i$. If we do not want to select the item we just bid $\pi_i=0$. We rewrite the problem by replacing our policy $\pi$ with equivalent selection variables $x_i \in \{0,1\}$ and have:
\begin{align}
\label{eq: kpi-opt-trad} \tag{K}
\max_{x \in \{0,1\}} \sum_i v_i x_i \st \sum_i b_i x_i \leq B
\end{align}

\subsection{Near Optimality of the Linear Bid}\label{sec:dual}
In this section we build an intuition for why the connection between the second-price knapsack problem and the traditional knapsack problem might be expected. This is done through an analysis of the offline problem.

We make use of two assumptions, that are observed to be reasonable based on an empirical review of the iPinYou data (see Section \ref{sec:empiricalresults}).
\begin{assumption} [{Value-to-Price Uniqueness}] \label{ass:unique} The ratios $\frac{v_i}{b_i}$ are unique for our selection of ads $I$. 
\end{assumption}
\begin{assumption} [Low-Individual-Impact] \label{ass:low} For each ad, $b_{j} \ll B$ and $v_{j} \ll Z_{IP}(B)$.
\end{assumption}

From these assumptions, the first result for the second-price knapsack problem follows: there exists an implementable bidding policy $\pi$ that recovers a near optimal selection of ads for the offline setting, where the bid amount is only a function of the same information that is available when ads arrive online. In particular, this near-optimal bid amount is \textit{not} a function of the price that must be paid in order to win the auction!

\begin{proposition}\label{thm:DT}
	Consider the optimization problem (\ref{eq: kpi-opt}): for any set of ads $\{i \in I\}$ with values $v_i \leq v_{\max}$, we seek a bidding policy $ \pi $ to maximize total value of ads purchased subject to our budget $B$.
	Then there exists some constant $\lambda^* \in \mathbb{R}^+$ such that the linear bidding policy $\pi$ which bids an amount \[ \pi_i = \frac{v_i}{\lambda^*} \] 
	yields a feasible set of purchased ads with a total value within $v_{\max}$ of the optimal value.
\end{proposition}

This is very similar to a recent result by \citet{conitzer2017multiplicative}. However our work, which was done independently, has a couple distinctions which are valuable to this paper. First, the \citet{conitzer2017multiplicative} result proves optimality of the linear bid by assuming that the bidder can arbitrarily choose to purchase any subset of the ads where the amount we bid $\pi_i$ equals the highest competitor bid, and requires some a priori knowledge about how many ads this will apply to. In contrast, we directly show near-optimality without this a priori knowledge. The second difference is that their proof is a direct argument, while ours uses the dual problem to prove near-optimality. Because the dual interpretation is used in Sections \ref{sec:onlineselection} and \ref{sec:empiricalresults}, this version of the proof gives some additional insight as it relates to the main results in Section \ref{sec:onlineselection}.

An overview of the proof is as follows, let us again consider the traditional knapsack problem in (\ref{eq: kpi-opt-trad}). The linear programming relaxation (\ref{eq:LP}) can then be written as:
\begin{align}
\max_{x_i \in [0,1]}   & \sum_i v_i x_i \st \sum_i b_i x_i \leq B \tag{K-LP}\label{eq:LP}
\end{align}
The dual problem to (\ref{eq:LP}) is:
\begin{align*}
&\min_{\lambda, z_i \geq 0} \lambda B +  \sum_i z_i \st \forall i:  z_i \geq v_i - \lambda b_i
\label{eq:dual} \tag{K-Dual}
 \\
&=\min_{\lambda \geq 0} \lambda B +  \sum_i (v_i - \lambda b_i)^+ =\min_{m} \frac{v_m}{b_m} B + \sum_{i=1}^{m} (v_i - \frac{v_m}{b_m}b_i)
\end{align*}
{where ads are such that $ i \leq j \Rightarrow {v_i}/{b_i} \geq {v_j}/{b_j} $.}
Dual prices allow us to solve (\ref{eq:LP}) exactly: we select an item if and only if $\frac{v_i}{b_i}$ is above a certain threshold parameter $\lambda^*$ \citep[see][]{fisher1981lagrangian, kellerer2004introduction, bertsekas1995dynamic}. We then demonstrate that the optimality gap between (\ref{eq:LP}) and (\ref{eq: kpi-opt-trad}) is bounded by a small amount because $b_i \ll B$. Finally, we turn offline selection into an implementable online bidding policy for the second-price knapsack problem (that is not a function of $b_i$).

A full proof of Proposition \ref{thm:DT} is given in Appendix \ref{sec:PropositionProof}.

\subsection{Adapting Online Knapsack Algorithms}\label{sec:onlineselection}
In this section we present our main results (Theorems \ref{thm:dualOnline} and \ref{thm:primalOnline}): methods for adapting online knapsack selection algorithms on finite time horizon $ t \in \{1,...,T\} $, where we select ads based on some function of $(v_t, b_t)$ at time $ t $ where the paying price is known, to the second-price online knapsack setting, where we must make a bid without knowledge of $b_t$ a priori. 
In essence, these methods provide a generalization of selection policies, ill-defined when the paying price is unknown, to implementable bidding policies, which are now suited for the second-price problem.\\
We present two such methods. The first is a method for deterministic selection policies, and applies widely to dual online knapsack algorithms, and the second is a method for randomized selection policies, and applies widely to primal online knapsack algorithms. We give examples for each method, including a one-shot learning algorithm, that creates an online, near-optimal bidding policy under the random permutation setting. 

We say that a deterministic algorithm has \textbf{competitive ratio} $\alpha$ for the online problem if:
\[ \mathbb{E}_{I} \big[ \frac{ALG(I)}{OPT(I)} \big] \geq \alpha\]
where $OPT(I)$ denotes the performance obtained by the offline optimum  on a random instance $I$, $ALG(I)$ is the performance of the online algorithm, and the expectation is taken over the randomness in online instances.

Because the online knapsack algorithms of interest to us rely on the random permutation assumption, we explore what this requires in the second-price knapsack problem. We assume a stable setting, where our competitors have consistent bidding strategies, the auction platform keeps consistent reserve prices, and ad opportunities arrive in a random order. From these assumptions we recover the \textit{random permutation assumption} on an expanded set of ads, including some synthetic \textit{zero-value ads}. This property states that for a set of future ads, they are equally likely to arrive in any order.
These assumptions directly imply the following lemma.

\begin{assumption}\label{ass:stablearrival}
Conditioned on $n$ ads arriving over some period of time, all permutations of the arrival order are equally likely. 
\end{assumption}

\begin{assumption}\label{ass:stableprice}
	Competitors have consistent bidding strategies and the auction platform keeps consistent reserve pricing methods over time. That is, given ad $i$ with specific contextual parameters (giving us value $v_i$), the competitors bid the same amount for the ad, and the auction platform sets the same reserve price for us, regardless of when it arrives. As a result, paying price $b_i$ is the same for us if we win the ad, regardless of which week the ad arrives in.
\end{assumption}

\begin{lemma}[Random Permutation]\label{lm:randomsample}
	Under Assumptions \ref{ass:stablearrival} and \ref{ass:stableprice}, we recover the random permutation assumption.
\end{lemma}

\proof{}
From assumption \ref{ass:stablearrival}, any arrival of ads, as measured by their values $v_i$, are equally likely. From assumption \ref{ass:stableprice}, the paying prices $b_i$ remain the same for all ads, regardless of the order they arrive in. As a result any arrival of ads, as measured by both $v_i$ and $b_i$ are equally likely, which is the random permutation property.
\endproof
The random permutation assumption is technical, only necessary to recover the same theoretical guarantees as online knapsack algorithms which use a variation of the assumption. In practical settings where the assumption can be violated, our methodology can still be applied.

We continue to our main results: methods for taking a wide class of online knapsack algorithms, where we select the ad at time $ t $ based on some function of $v_t$ and $b_t$, and turning them into online bidding strategies for the second-price knapsack problem, where we need to submit a bid without prior knowledge of the price $b_t$. 

For the remainder of the section, we let $B_t$ denote the remaining budget at time $t$, and denote $\mathcal{H}_t$ the history available to the bidder at time $t$ (including the values of previous ads, the maximum of our competitors' bids, the reserve price, and the auction outcome).

\subsubsection{Deterministic Algorithms}
The first method summarized in Theorem \ref{thm:dualOnline} holds for deterministic online knapsack selection algorithms, and follows a similar reasoning to Lemma \ref{lm:linear bid} from the appendix. To the best of our knowledge, this Theorem can be applied to all dual-based online knapsack algorithms.

\begin{theorem}\label{thm:dualOnline}
Consider an online knapsack algorithm, where at time step $t$ we select an ad if $g_t(v_t,b_t, B_t, \mathcal{H}_t)\leq 0$, for some continuous function $g_t$ which is monotonically increasing with respect to $b_t$ and satisfies $g_t(b_t=0) \leq 0$ and ${\lim}_{b_t \to \infty}g_t(.) > 0$. 
\\In the second price setting, we can recover exactly this set of ads without a priori knowledge of $b_t$, and therefore achieve the same competitive ratio, by bidding: 
\[ b'_t = \sup \{b \mid g_t(v_t,b, B_t, \mathcal{H}_t)\leq 0\} \]
Further if $ g_t $ implements a budget feasible selection algorithm then our bidding policy is also budget feasible.
\end{theorem}
\proof{}
We prove that this algorithm recovers all ads where\\$g_t(v_t, b_t, B_t, \mathcal{H}_t) \leq 0$, and no ads where $g_t(v_t,b_t, B_t, \mathcal{H}_t) > 0$. If $g_t(v_t,b_t, B_t, \mathcal{H}_t) \leq 0$, then because $g_t$ is monotonically increasing with respect to $b_t$, we have that $b'_t \geq b_t$, so we win the auction by bidding $b'_t$. Similarly if $g_t(v_t,b_t, B_t, \mathcal{H}_t) > 0$, then because $g_t$ is monotonically increasing with respect to $b_t$, we have that $b'_t < b_t$, so we do not win the auction by bidding $b'_t$. 

Because $b'_t = \text{sup}\{b \mid g_t(v_t,b, B_t, \mathcal{H}_t)\leq 0\}$ is not dependent on the a priori unknown price $b_t$, this is an implementable bidding strategy in the online second price knapsack problem. 
Further given feasibility of the selection policy, then we have $ b>B_t \Rightarrow g_t(.,b,B_t,.) > 0 $, so our bids verify $ b'_t \leq B_t $ which guarantees feasibility of our bidding policy.
\endproof

We now give an example of how Theorem \ref{thm:dualOnline} can be applied. We consider existing work on a one-shot learning for the online knapsack by \citet{agrawal2014dynamic}. Using this, we give a bidding policy for the online second price knapsack problem, that has a competitive ratio of $1-6\epsilon$ under the random permutation setting. While there are improved results upon which Theorem \ref{thm:dualOnline} could be applied \citep[e.g.,][]{agrawal2015fast}, we think the given example is a very intuitive approach, shares similarities to Proposition \ref{thm:DT}, gives a good introduction to the process of adapting deterministic online knapsack algorithms, and offers a low latency making it practical for DSPs to implement.

\begin{example}[\citet{agrawal2014dynamic}]\label{cor:Gen-RTB}
	
	Let $B$ denote total budget and let $b_{\max}$ be an upper bound on the paying price for any ad. Select a training ratio $\epsilon \in (0,1)$, s.t. $\epsilon$ satisfies:
	\begin{equation} \label{eq:budg_ag}
	B \geq \frac{6 \times b_{\max} \times  {\log}(n/ \epsilon)}{\epsilon^3}
	\end{equation}
	For the first fraction $\epsilon$ of ads, we record the highest bid $b_i$ for each ad, after the auction is completed. This is what we would have had to bid (and pay) in order to win the ad. Determine the optimal dual price solution $\lambda^*$ to (\ref{eq:online_trainLP}) given below:			
	\begin{align}
\max_{x_t \in [0,1]} \sum_{t \leq \epsilon T} v_t x_t \st \sum_t b_t x_t &\leq (1-\epsilon)\epsilon B \tag{OLA-LP}\label{eq:online_trainLP}
	\end{align}
	We then apply Theorem \ref{thm:dualOnline}, and turn the selection policy of \citet{agrawal2014dynamic} into a bidding policy for the remaining fraction $1-\epsilon$ of ads. For these remaining ads, we bid $\pi_t = \min\{\frac{v_t}{\lambda^*},B'\}$ for ads $t > \epsilon T$, where $B'$ is our remaining budget. The bidding policy $\pi$ is feasible by definition in the second-price setting. 
	From Theorem \ref{thm:dualOnline}, we recover the same set of ads as would the algorithm from \citet{agrawal2014dynamic} in the traditional knapsack setting, therefore recovering the same competitive ratio of $1-6\epsilon$. Given the offline optimum remains unchanged under both settings, our adaptation to the problem where the paying prices are unknown enjoys the same competitive ratio of $1-6\epsilon$ relative to the optimal bidding policy. The methodology in Theorem \ref{thm:dualOnline} allowed to adapt the existing selection policy when paying prices are known to our bidding policy under the unknown second price.

	Contextually, since our budget is very large relative to ad prices, this process could be applied for a very small $\epsilon$, and would give a very strong competitive bound. For clarity we now present Example \ref{cor:Gen-RTB} in Algorithm \ref{alg:OSLA} as pseudo-code.

\begin{algorithm}[h]
	\caption{One-Shot Learning Algorithm}\label{alg:OSLA}
	\begin{algorithmic}[1]
		\Procedure{Near-Optimal Bidding Strategy}{$\epsilon, T$} \\ \Comment{$ t $ denotes time, $ T $ the number of periods, $ \epsilon $ s.t. (\ref{eq:budg_ag}) and $ \epsilon T \in \mathbb{N} $.}
		
		\While{$t < \epsilon T$}
		\State Record ad value and paying price. Continue.
		\EndWhile
		
		\State Determine optimal dual parameters $\lambda^*$ on impressions in first $\epsilon T$ weeks from (\ref{eq:online_trainLP}), as in Lemma \ref{lm: CSC}.
		
		\While{$t < T$}
		\State For impression $t$, bid $\min\{\frac{v_t}{\lambda^*},B'\}$ where $B'$ is the left-over budget
		\State If the auction is won, collect value and update budget
		\EndWhile
		\EndProcedure
	\end{algorithmic}
\end{algorithm}

\end{example}

\subsubsection{Randomized Algorithms}

The second method, summarized in Theorem \ref{thm:primalOnline}, holds for randomized online knapsack selection algorithms. This Theorem can be applied to a variety of primal online knapsack algorithms, to adapt them into randomized bidding policies for the second-price knapsack problem.

\begin{theorem}\label{thm:primalOnline}
Consider an online knapsack algorithm, where at time step $t$ we select an ad with probability $p_t = p_t(v_t, b_t, B_t, \mathcal{H}_t)$, for some (right-)continuous function $p_t$ which is decreasing with respect to $b_t$, and satisfies $\lim_{b_t \to 0}p_t(.) \to 1$ and ${\lim}_{b_t \to \infty}p_t(.) \to 0$. By defining $F_t(b)=1-p_t(v_t,b, B_t, \mathcal{H}_t)$, we create a CDF with domain $b \in [0,\infty]$. \\
In the second price setting, we can purchase ads with the same probability as $p_t$, without a priori knowledge of $b_t$, and therefore achieve the same competitive ratio, by bidding:
\[b'_t = \sup\{b \mid F_t(b) \leq u_t \} \text{ where } u_t \sim \text{Unif }[0,1], i.i.d \]
Further if $ p_t $ implements a budget feasible selection algorithm with probability 1, then our bidding policy is also budget feasible.
\end{theorem}
 
\proof{}
We first show that $F_t$ is well defined. Because $p_t$ is decreasing in $ b $, and satisfies $\text{lim}_{b^+ \to 0}p_t(.) \to 1$ and $\text{lim}_{b \to \infty}p_t(.) \to 0$, we have that $F_t(b)=1-p_t(v_t,b, B_t, \mathcal{H}_t)$ is increasing, with domain $b \in [0,\infty]$, with $\text{lim}_{b^+ \to 0}F_t(b) \to 0$ and $\text{lim}_{b \to \infty}F_t(b) \to 1$. Therefore $F_t$ defines a (right-)continuous CDF for some distribution. 

Next, we show that randomly drawing $u \in \text{U}[0,1]$, and then bidding $b'_t = \sup\{b \mid F_t(b) \leq u \}$, will recover ad $t$ with probability $p_t(v_t,b_t, B_t, \mathcal{H}_t)$. In the second price setting, we recover ad $t$ if and only if we bid $b'_t \geq b_t$. 
This happens when $u \geq F_t(b_t)$, which occurs with probability $1-F_t(b_t)$. From the definition of $F_t$, we have $1-F_t(b_t)=p_t(v_t,b_t, B_t, \mathcal{H}_t)$, so we recover the ad with the initial desired probability. 

Because $b'_t = \sup\{b \mid F_t(b) \leq u \}$ is not dependent on the a priori unknown price $b_t$, this is an implementable randomized bidding strategy in the online second price knapsack problem. 
Further given feasibility of the randomized selection policy, then we have $ b>B_t \Rightarrow p_t(.,b,B_t,.) = 0 \Rightarrow F_t(b) = 1 > u $ a.s. Thus our bids verify $ b'_t \leq B_t $ a.s., which guarantees feasibility of our randomized bidding policy.
\endproof

Next, we give an example of how Theorem \ref{thm:primalOnline} can be applied, to create a bidding strategy for the second-price knapsack problem with a strong competitive ratio in the stable setting.

\begin{example}[\citet{kesselheim2014primal}]\label{cor:primalbeatsdual-RTB}
	
	Let $B$ denote total budget and let $b_{\max}$ be an upper bound on the paying price for any ad. 
	Similarly, given a scaling factor $ f $ and a set of observed ads $ S \subset [1,T]$, we denote $ \mathcal{P}(f,S) $ the set of feasible solutions to the linear relaxation of (\ref{eq: kpi-opt-trad}) where the budget $ B $ is scaled by $ f $ and only ads from $ S $ can be selected. 
	Formally this is the set $ \{x \mid 0 \leq x_S \leq 1,\: \sum_{i \in S} x_i \leq f B, \andd x_{\bar{S}} = 0 \} $.
	
	For every ad arrival $ t $, we solve a fractional scaled ad selection for the knapsack problem, where in particular the scaling factor is $ \frac{t}{T} $. We then interpret the fractional selection as a randomized selection policy, which can be readily implemented as a bidding policy in the second price setting using Theorem \ref{thm:primalOnline}. 
	In particular at ad arrival $ t $, we let $ \tilde{x}(b_t) $ the function of optimal solutions to the scaled problem 
		\begin{align}
\max_{x\in \mathcal{P}(t/T, [1,t])} \sum_{t'\leq t} x_{t'} v_{t'}  \tag{PLA-LP}\label{eq:ex2_solution}
	\end{align}
		with paying price $ b_t $ and we define the CDF $ F_t $ accordingly. Then we sample $ u\sim Unif[0,1] $, and bid $ \min(B_t, \sup\{b \mid F_t(b) \leq u \}) = \min(B_t, \sup\{b \mid \tilde{x}(b) \geq 1 - u \}) $ 
	where $ B_t $ is the leftover budget.
	
	Consider a modified version of our strategy. We first bid regardless of our leftover budget, but only carry out the selection if the paying price is smaller than our leftover budget. That is we bid $ \textbf{1}_{\sup\{b \mid F_t(b) \leq u \} \leq B_t}\times\sup\{b \mid F_t(b) \leq u \} $. This modified bidding policy realizes the same probabilistic selection policy from the randomized rounding procedure in \citet{kesselheim2014primal}. 
	Therefore Lemma 2 and subsequently Theorem 3 from \cite{kesselheim2014primal} hold for our proposed bidding policy. 
	
	Finally our policy achieves the competitive ratio of $ 1 - 45\sqrt{\frac{b_{\max}}{B}} $. Equivalently $ \forall \epsilon >0 $, for $ B \geq \frac{b_{\max} }{\epsilon^2} $, the randomized algorithm achieves an expected competitive ratio of $ 1-45\epsilon $ relative to the optimal bidding policy over the entire set of ads $[1,T]$.
		
\end{example}

Again in our context this process could be applied for small $\epsilon$, and would give a very strong competitive ratio. For clarity we now present Example \ref{cor:primalbeatsdual-RTB} in Algorithm \ref{alg:primal-alg-rtb} as pseudo-code.

\begin{algorithm}[h]
	\caption{Primal based randomized algorithm}\label{alg:primal-alg-rtb}
	\begin{algorithmic}[1]
		\Procedure{Near-Optimal Bidding Strategy}{$\epsilon, T$} \\ \Comment{$ t $ denotes time, $ S_t $ the observed ads, $ T $ the number of periods.}
	
		\For{$t < T$}		
		\State Let $ \tilde{x}_t $ the optimal solution to the scaled problem (\ref{eq:ex2_solution}) by a factor $ \frac{t}{T}$ on ads $ S_t $,
		if the paying price were $ b_t $.
		\State Sample $ u\sim Unif[0,1] $. Bid $  \min(B_t, \sup\{b \mid \tilde{x}_t(b) \geq 1 - u \}) $.
		\State If the auction is won, collect value and update budget
		\EndFor
		
		\EndProcedure
	\end{algorithmic}
\end{algorithm}

We conclude this section by pointing that our theorems generalize to deterministic auctions, where for every auction, there exists a value $ b $, such that the bidder wins the auction if they bid at least $ b $ and pay exactly $ b $. The proofs can be directly adapted using this simple property.

\section{Empirical Results}\label{sec:empiricalresults}
In this section we use the iPinYou dataset to give numerical support to our work. Assigning $v_i$ as the expected click-through rate per advertiser for that impression, we measure the efficiency of Algorithm \ref{alg:OSLA} and show we are able to recover near-optimal ad bundles in an online setting. These results are robust to changes in budget, and are consistent for all advertisers, which highlights the effectiveness of our methods. We also compare our results against adaptive pacing.

\subsection{iPinYou Dataset}
Until recently, academics were limited in studying the application of RTB strategies since bidding data is generally kept secretive. Fortunately in 2013, iPinYou Information Technologies Co., Ltd (iPinYou), the largest DSP in China, began a competition for RTB algorithms and released three \textit{seasons} of data for a small number of advertisers. Each season corresponds to one week of data, with the entire release totaling 35GB. To the best of our knowledge, this is the first publicly available RTB dataset. 
	
\paragraph{Data Format}
For the competition, iPinYou released information for different types of exchange activity: bids, impressions, clicks, and conversions. Combined, these datasets capture most of the relevant data from an auction: 1) The contextual ad features which are sent along with bid requests (ad slot parameters, viewer demographics), 2) The winning bid amount and the paying price (which we refer to as the market price), and 3) The user feedback, i.e. clicks and conversions on the won impression. The dataset variables and advertisers also vary by season, so we chose to focus our numerical testing on season three of the data, which included advertiser and user IDs.

The summarized features for the data and a full description are provided in Tables \ref{table: Log} \& \ref{table: Log-appendix}.

\begin{table}[h]
	\caption{The summarized log data format for an impression and their description.}
	\label{table: Log}	\centering
	\scalebox{0.8}{\begin{tabular}{cl}
		\toprule
Feature & Description\\
		\midrule 
iPinYou ID 	& A unique identifier for every bid in an auction. \\
Timestamp &  Date of the auction.\\
Log type & Outcome of the ad - whether the user clicked or purchased.\\
Bidding price & The amount bid by the advertiser.\\
Paying price & The amount paid by the winner of the auction.\\
Advertiser ID & Information concerning the advertiser.\\
		\bottomrule
	\end{tabular}}
\end{table}

\begin{table}[h]
	\centering	
	\caption{The log data format for an impression}	
	\scalebox{0.8}{\begin{tabular}{c|l}
		\toprule
		Col \# & Feature\\
		\midrule 
		1 & Bid ID 	\\
		2 & Timestamp\\
		3 & Log type \\
		4 & iPinYou ID \\
		5 & User-Agent\\
		6 & IP\\
		7 & Region\\
		8 & City \\
		\bottomrule
	\end{tabular}
	\quad
	\begin{tabular}{c|l}
		\toprule
		\\
		\midrule
		9 & Ad Exchange\\
		10& Domain \\
		11&URL\\
		12& Anonymous URL ID\\
		13& Ad slot Id\\
		14& Ad slot width\\
		15& Ad slot height\\
		16& Ad slot visibility\\
		\bottomrule
	\end{tabular}
	\quad
	\begin{tabular}{c|l}
		\toprule
		\\
		\midrule
		17& Ad slot format\\
		18& Ad slot floor price\\
		19& Creative ID\\
		20& Bidding price\\
		21&Paying price\\
		22& Key page URL\\
		23& Advertiser ID\\
		24& User Tags\\
		\bottomrule
	\end{tabular}}
	\label{table: Log-appendix}
\end{table}

\begin{figure}[h]
	\caption{Effective CPC against different features\label{fig:ecpc}}	
	\begin{center}
		\includegraphics[width=0.45\textwidth]{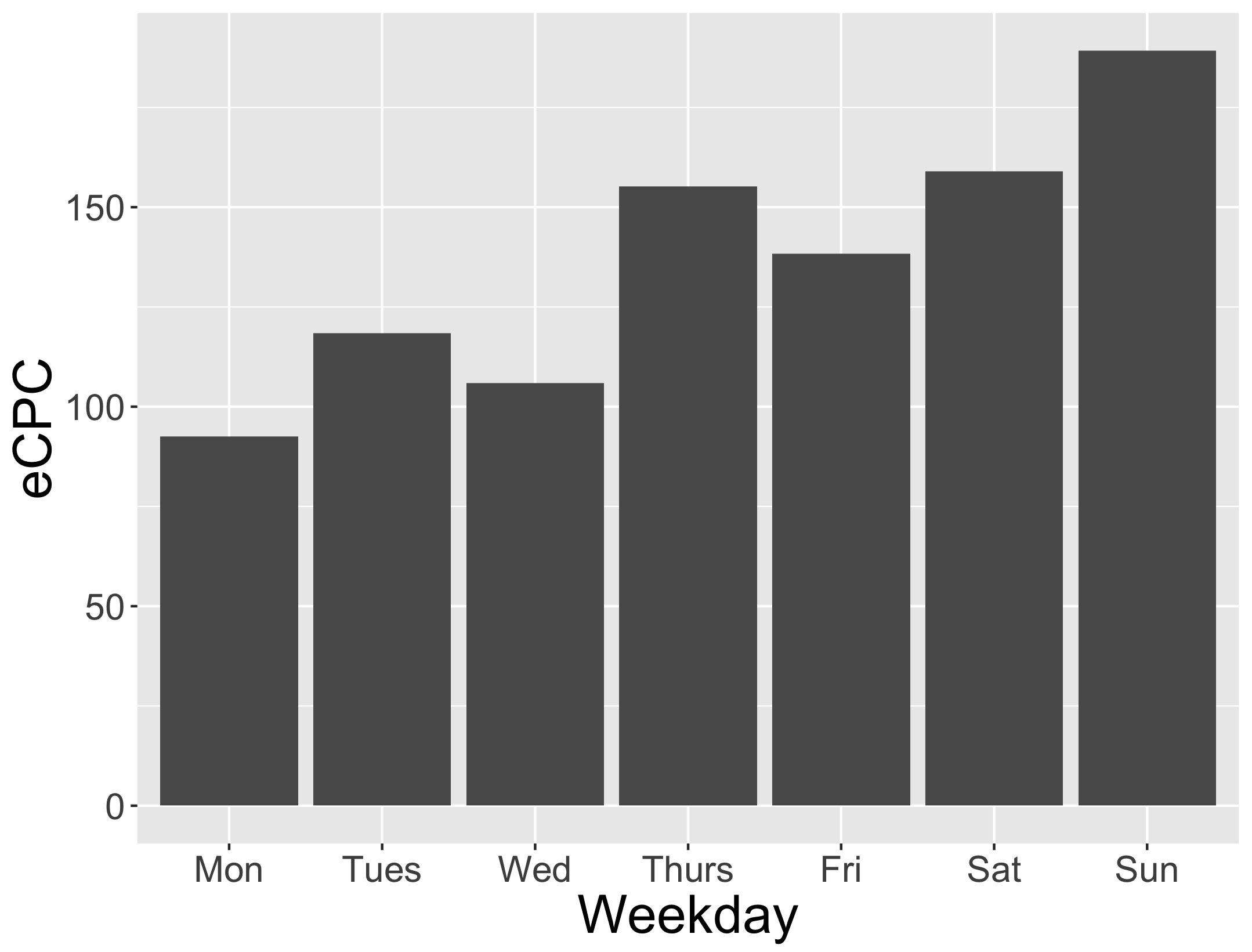}
		\includegraphics[width=0.45\textwidth]{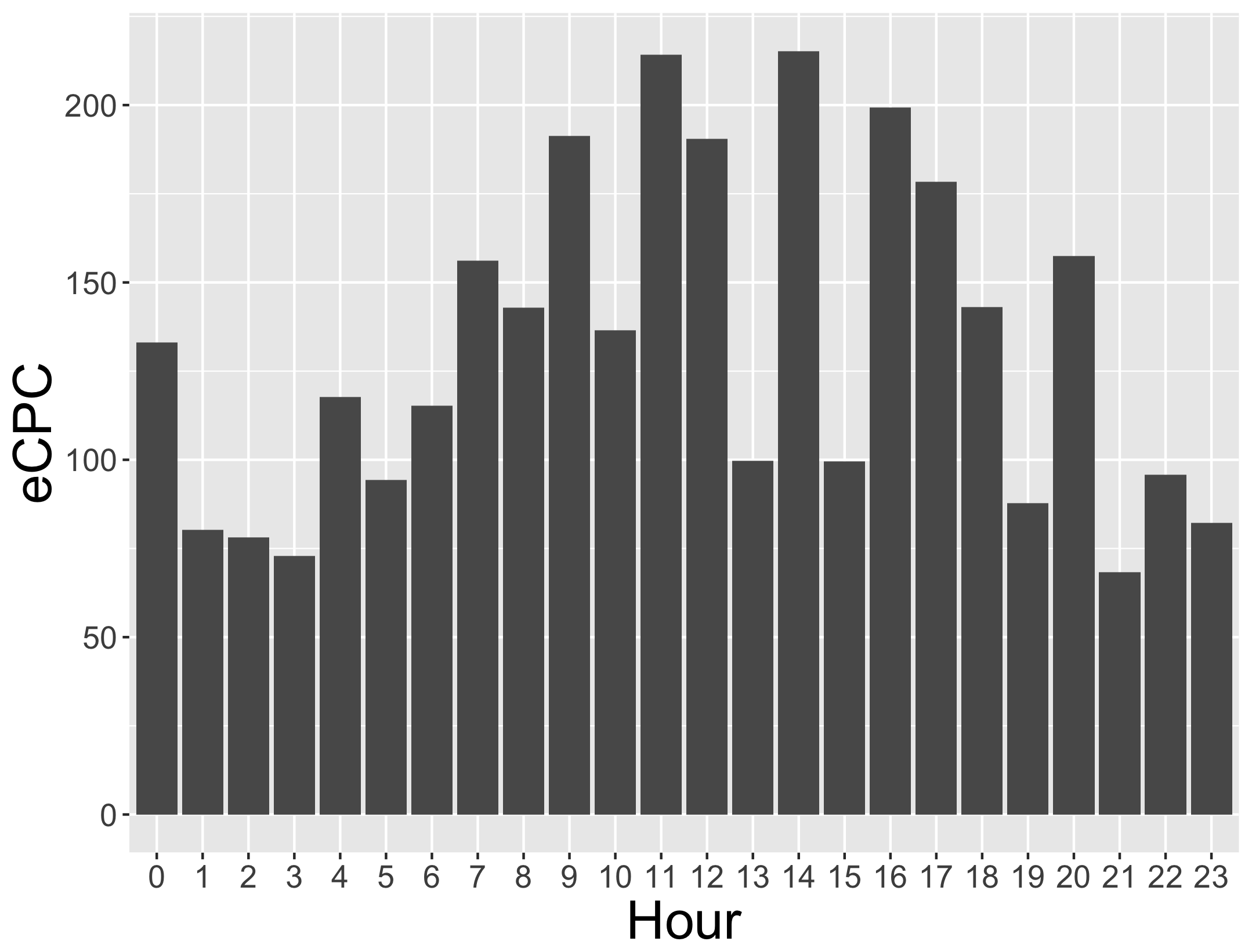}
	\end{center}
\end{figure}

\paragraph{Data Summary Statistics}
Season three contains information about five advertisers in different fields: Chinese vertical e-commerce, software, international e-commerce, oil and tires. All advertisers have click-through rate CTR on their won auctions inferior to 0.1\%. Although the cost for achieving an impression is similar across advertisers ($\approx$80 RMB per thousand impressions), the expected cost per click differs by advertiser. Further the conversion rate {CVR} varies also. From a review of the data it was clear the extent to which features affected the CTR, with the influence of a given feature varying for each advertiser. This motivated modeling a different predicted click-through rate \textbf{(pCTR)} estimator $pCTR$ for each advertiser, to account different customer feedback trends depending on the day and time. From the logs, we also review the variations of cost-per-click (CPC) compared to individual features. We plot relevant CPC variations against different features in Figure \ref{fig:ecpc}.

This brief review of the iPinYou dataset suggests there exist opportunities for more efficient bidding strategies among each advertiser. In addition we also report the basic statistics from the third season of data (June 6-12) in Table \ref{table: data}.

\begin{table}[h]
	\centering\caption{Season Three - Data Summary \label{table: data}}
		\scalebox{0.75}{\begin{tabular}{c|cccccc}
			\toprule
			Advertiser ID & Impressions & Clicks & Tot. Bid & Tot. Cost & CTR & eCPC\\
			\midrule 
			11458 & 	 3,083,055 &  	 2,452 &  	 924,916,500 &  	 212,400,191 &  	 0.080\% &  	 86.62 \\  
3358 & 	 1,742,104 &  	 1,350 &  	 405,466,041 &  	 160,943,087 &  	 0.077\% &  	 119.22 \\  
3386 & 	 2,847,802 &  	 2,068 &  	 854,340,600 &  	 219,066,895 &  	 0.073\% &  	 105.93 \\  
3427 & 	 2,593,765 &  	 1,922 &  	 612,619,071 &  	 210,239,915 &  	 0.074\% &  	 109.39 \\  
3476 & 	 1,970,360 &  	 1,027 &  	 488,187,425 &  	 156,088,484 &  	 0.052\% &  	 151.98 \\  
			\bottomrule
		\end{tabular}}
\end{table}
	
\subsection{Data Pre-Processing}
Starting with the iPinYou data shown in Table \ref{table: Log}, processing was performed to get the data in a usable format. We first deleted the small number of ad impressions for which there was missing data and duplicated to remove a small number of redundant rows. We extracted the weekday and hour feature from timestamps. From \textit{user agent} text, we extracted operating systems. We also split the column of the tag list of user interests into a large number of binary variables, representing user interests in all the specific categories. We also removed unique features or nearly unique such as Bid ID, Log Type, iPinYou ID, URL, Anonymous URL ID and Key Page URL that we could not train our predictors on.

\subsection{Experimental Setup}
Success in our empirical work is measured with respect to some known values for each ad. In practice, advertisers value success based on their click-through-rate (CTR) and conversion rate. Each advertiser values the balance between clicks and conversions differently, which is captured in the general literature as the Key Performance Indicator (\textbf{KPI}), defined as a linear combination: 
\[ \text{KPI} = \mathbbm{1}_\text{clicks } + \alpha \times \mathbbm{1}_\text{conversions } \] 
where the factor $ \alpha $ is advertiser specific. However as seen in Table \ref{table: data}, the small number of conversions presented an issue and we made the decision to set $\alpha = 0$ for our testing, meaning $KPI = pCTR$. 

To evaluate the threshold strategy, we therefore assign ad values as follows: we create a $pCTR$ estimator for each advertiser, and set the value of each ad $v_i$ equal to the $pCTR$ estimate for the ad. To calculate $pCTR$ for each ad we split most categorical features into multiple columns of \{0,1\}, such as the long-term interests and the demographic fields. A single ad is represented by vector of over 50 features. We then relied on logistic regressions from \textit{R's GLM} package to predict the probability of a click. While this gave us a good estimator and was sufficient for our purposes, in practice the $pCTR$ prediction is an essential piece of any bidding strategy and advertisers would be well-served by devoting considerable resources to creating the best prediction model possible. In contrast, our priority is just evaluating the effectiveness of our algorithms given an assumed set of values $v_i$ for ads chosen reasonably. 

A common difficulty in the experimental setup for RTB research comes from the fact that we only possess detailed ad information for auctions that the advertiser won. Thus in our testing phase, we cannot let an advertiser buy an advert they did not win. We approached this the same manner as previous researchers, and we design a bidding system where each advertiser is only shown auctions they won from the data, but we allow each advertiser a budget only a fraction of the amount they actually spent (otherwise with full budget, the optimal strategy would always be to win every ad they previously won at the same price, which is made possible through bidding your remaining budget at every step because of the second-price auction mechanism). We chose to evaluate several fractional budget amounts, and we set the budget for each advertiser as $\frac{1}{2}$, $\frac{1}{4}$, $\frac{1}{8}$, and $\frac{1}{16}$ of the original total they actually spent. 

Let us specifically review the testing process for a fixed advertiser under a budget limit. At each time step:
\begin{enumerate}
\item We draw an ad randomly from our test set (that the current advertiser won from our data) and we pass it as a bid request.
\item The bidding strategy computes a bid amount for this contextual request, which does not exceed the remaining budget.
\item If the bid is higher than the paying price, then the advertiser wins the auction. The paying price is subtracted from the remaining budget and the value $v_i$ is added to the objective.
\end{enumerate}

\subsection{Results}
In this section we present results of testing. To create our testing data, we randomly select one million records per advertiser, for each of the five advertisers in the third season of the iPinYou data. We set our values $v_i$ for each advertiser, using a logistic regression for $pCTR$ on the entire testing data set. A summary of the testing data, including the total $pCTR$ predictions are given in Table \ref{table: seasonthreedata}.

\begin{table}[h]
	\centering\caption{Season Three Subset for Online Testing - Data Summary \label{table: seasonthreedata}}
		\scalebox{0.75}{\begin{tabular}{c|cccccc}
			\toprule
			Advertiser ID & Impressions & Clicks & Tot. Bid & Tot. Cost & CTR & eCPC\\
			\midrule 
			1458 & 	 1,000,000 &  	 800 &  	 300,000,000 &  	 68,884,867 &  	 0.080\% &  	 86.11 \\  
3358 & 	 1,000,000 &  	 798 &  	 232,746,786 &  	 92,410,850 &  	 0.080\% &  	 115.80 \\  
3386 & 	 1,000,000 &  	 732 &  	 300,000,000 &  	 76,918,986 &  	 0.073\% &  	 105.08 \\  
3427 & 	 1,000,000 &  	 742 &  	 236,188,992 &  	 81,096,832 &  	 0.074\% &  	 109.29 \\  
3476 & 	 1,000,000 &  	 515 &  	 247,763,552 &  	 79,181,869 &  	 0.052\% &  	 153.75 \\  
			\bottomrule
		\end{tabular}}
\end{table}

The next step was benchmarking the optimal bundle that could be achieved for a given budget in the test dataset, in the offline setting. We created benchmarks for each of the five advertisers, and each choice of budget ($\frac{1}{2}$, $\frac{1}{4}$, $\frac{1}{8}$, and $\frac{1}{16}$).
		
After benchmarking we tested the online version of Algorithm \ref{alg:OSLA} which follows from the methodology in Theorem \ref{thm:dualOnline}. 
To do this we determined the optimal threshold $\hat{\lambda}$ for each budget choice on the $\epsilon = 1.0\%$. During this training period we bid our value, and deplete the entire training budget. We then took the remaining data, and applied the dual threshold bidding with that same constant $\hat{\lambda}$ in the online setting. We recorded the total value $v_i = pCTR$ and actual clicks for the selection of ads. This is done for ten random permutations of the data\footnote{Because the data includes a very large number of ad auctions, and the value of each ad is small relative to the total, the results are extremely robust to random permutations. The standard deviation of results in the tables is less than 0.1\%.}. The test results are summarized in Table \ref{table: seasonthreetesting}.

	\begin{table}[ht!]
	\centering
			\caption{Online Algorithm \ref{alg:OSLA} Results on Season Three Testing. \label{table: seasonthreetesting}}
	\subfloat[][Budget 1/2 Results]{
		\scalebox{0.68}{\begin{tabular}{c|cccc}
			\toprule
			Advertiser ID & pCTR & \% of Opt. Bundle pCTR & Actual Clicks & \% of Opt. Bundle Clicks\\
			\midrule 
			1458 &	 756 &  	98.3\% & 	 754 &  	98.3\%     \\
3358 &	 741 &  	97.5\% & 	 728 &  	98.1\%     \\
3386 &	 587 &  	95.5\% & 	 490 &  	94\%     \\
3427 &	 686 &  	97.5\% & 	 620 &  	96.1\%     \\
3476 &	 482 &  	97.6\% & 	 426 &  	97\%     \\
Total &	 3,253 &  	97.3\% & 	 3,018 &  	96.9\%     \\
			\bottomrule
	\end{tabular}}}
	\qquad

	\subfloat[][Budget 1/4 Results]{
		\scalebox{0.68}{\begin{tabular}{c|cccc}
			\toprule
			Advertiser ID & pCTR & \% of Opt. Bundle pCTR & Actual Clicks & \% of Opt. Bundle Clicks\\
			\midrule 
			1458 &	 712 &  	98.5\% & 	 702 &  	98\%     \\
3358 &	 675 &  	97.3\% & 	 630 &  	97.7\%     \\
3386 &	 456 &  	94.9\% & 	 339 &  	93.6\%     \\
3427 &	 626 &  	97.7\% & 	 575 &  	97\%     \\
3476 &	 426 &  	97.5\% & 	 382 &  	97.4\%     \\
Total &	 2,894 &  	97.3\% & 	 2,628 &  	97\%     \\
			\bottomrule
	\end{tabular}}}
	\qquad

	\subfloat[][Budget 1/8 Results]{
		\scalebox{0.68}{\begin{tabular}{c|cccc}
			\toprule
			Advertiser ID & pCTR & \% of Opt. Bundle pCTR & Actual Clicks & \% of Opt. Bundle Clicks\\
			\midrule 
1458 &	 669 &  	98.4\% & 	 653 &  	98.3\%     \\
3358 &	 618 &  	97.6\% & 	 578 &  	97.8\%     \\
3386 &	 355 &  	95.7\% & 	 236 &  	94.8\%     \\
3427 &	 572 &  	97.9\% & 	 528 &  	97.6\%     \\
3476 &	 362 &  	97.6\% & 	 335 &  	98.8\%     \\
Total &	 2,576 &  	97.6\% & 	 2,330 &  	97.7\%     \\
			\bottomrule
	\end{tabular}}}
	\qquad

	\subfloat[][Budget 1/16 Results]{
		\scalebox{0.68}{\begin{tabular}{c|cccc}
			\toprule
			Advertiser ID & pCTR & \% of Opt. Bundle pCTR & Actual Clicks & \% of Opt. Bundle Clicks\\
			\midrule 
1458 &	 638 &  	98.8\% & 	 619 &  	98.0\%     \\
3358 &	 574 &  	98.1\% & 	 534 &  	98.2\%     \\
3386 &	 272 &  	95.6\% & 	 184 &  	95.3\%     \\
3427 &	 523 &  	97.1\% & 	 496 &  	96.3\%     \\
3476 &	 299 &  	97.3\% & 	 287 &  	96.6\%     \\
Total &	 2,306 &  	97.7\% & 	 2,120 &  	97.2\%     \\
			\bottomrule
	\end{tabular}}}
\end{table}

We also compared our algorithm's performance to that of adaptive pacing, over the randomly permuted data. We apply adaptive pacing in accordance with the algorithm used to show its theoretical properties \citep{balseiro2017learning}. We begin by bidding our value while learning to be more selective over time, and use the recommended step size (that is, $\mu_0 = 0$ and $\epsilon = \frac{1}{\sqrt{T}}$). 
We chose this parameterization in order to provide a fair benchmark that does not make use of prior information, for either method \footnote{In fact, with perfect prior information (e.g., the offline setting), it can be seen that the optimal initialization of adaptive pacing is just $ \mu_0 = \frac{1}{\lambda^*} - 1 $  and $ \epsilon = 0 $ (i.e. no adaptiveness), which recovers the optimal bundle of ads from Proposition \ref{thm:DT}, and actually does not feature any 'pacing' as the bid multiplier is constant. Therefore considering the best implementation of adaptive pacing in hindsight, for all possible values of $ \mu_0$ and $ \epsilon$, is not a fair comparison.}.
For each strategy, we record the percentage of the optimal bundle's pCTR that was recovered, as well as the performance ratio (defined as the value of Algorithm \ref{alg:OSLA} to that of adaptive pacing). The results are recorded in Table \ref{table: adaptivepacing}.

	\begin{table}[ht!]
	\centering
			\caption{Algorithm \ref{alg:OSLA} and Adaptive Pacing Comparison on Season Three Testing \label{table: adaptivepacing}}
	\subfloat[][Budget 1/2 Percent of Optimal Bundle Recovered]{
		\scalebox{0.68} {\begin{tabular}{c|ccc}
			\toprule
			Advertiser ID & Algorithm 1 & Adaptive Pacing & Performance Ratio\\
			\midrule 
1458 &	98.3\% & 	99.5\% & 	98.8\% \\ 
3358 &	97.5\% & 	99.4\% & 	98.1\% \\ 
3386 &	95.5\% & 	99\% & 	96.5\% \\ 
3427 &	97.5\% & 	99.4\% & 	98.2\% \\ 
3476 &	97.6\% & 	99.6\% & 	98\% \\ 
Total &	97.3\% & 	99.4\% & 	98\% \\ 			\bottomrule
	\end{tabular}}
}\qquad
	\subfloat[][Budget 1/4 Percent of Optimal Bundle Recovered]{
		\scalebox{0.68}{\begin{tabular}{c|ccc}
			\toprule
			Advertiser ID & Algorithm 1 & Adaptive Pacing & Performance Ratio\\
			\midrule 
1458 &	98.5\% & 	97.8\% & 	100.7\% \\ 
3358 &	97.3\% & 	98.3\% & 	99\% \\ 
3386 &	94.9\% & 	96.5\% & 	98.4\% \\ 
3427 &	97.7\% & 	98\% & 	99.7\% \\ 
3476 &	97.5\% & 	98.1\% & 	99.3\% \\ 
Total &	97.3\% & 	97.8\% & 	99.5\% \\ 
			\bottomrule
	\end{tabular}}
}\qquad
	\subfloat[][Budget 1/8 Percent of Optimal Bundle Recovered]{
		\scalebox{0.68}{\begin{tabular}{c|ccc}
			\toprule
			Advertiser ID & Algorithm 1 & Adaptive Pacing & Performance Ratio\\
			\midrule 
1458 &	98.4\% & 	93.3\% & 	105.5\% \\ 
3358 &	97.6\% & 	95\% & 	102.7\% \\ 
3386 &	95.7\% & 	88.9\% & 	107.6\% \\ 
3427 &	97.9\% & 	93.9\% & 	104.2\% \\ 
3476 &	97.6\% & 	92.4\% & 	105.6\% \\ 
Total &	97.6\% & 	93.1\% & 	104.8\% \\ 			\bottomrule
	\end{tabular}}
}

	\subfloat[][Budget 1/16 Percent of Optimal Bundle Recovered]{
		\scalebox{0.68}{\begin{tabular}{c|ccc}
			\toprule
			Advertiser ID & Algorithm 1 & Adaptive Pacing & Performance Ratio\\
			\midrule 
1458 &	98.7\% & 	78\% & 	126.6\% \\ 
3358 &	98\% & 	83.7\% & 	117.1\% \\ 
3386 &	95.5\% & 	66.1\% & 	144.4\% \\ 
3427 &	97\% & 	78.5\% & 	123.6\% \\ 
3476 &	97.2\% & 	73.3\% & 	132.7\% \\ 
Total &	97.6\% & 	77.5\% & 	125.9\% \\ 			\bottomrule
	\end{tabular}}
}
\end{table}

The results are interesting. For advertisers with a large budget fraction of $\frac{1}{2}$ (enough to afford half the ads they are interested in), we see that adaptive pacing performs almost perfectly, which in fact validates our parameter tuning for their algorithm. 
In contrast, though, we see that the performance of adaptive pacing degrades significantly as advertisers become more selective. At budget of $\frac{1}{16}$, which represents a common level of selectivity among bidders in reality, adaptive pacing incurs significant performance degradation. This makes sense intuitively for selective bidders, as adaptive pacing's dual parameter search takes longer to converge to the optimum, and then oscillates more significantly around the optimum because of the step direction imbalance. At this budget fraction of $\frac{1}{16}$, the algorithm buys roughly that fraction of impressions, so about $\frac{15}{16}$ ad arrivals are not purchased and lead the algorithm to become slightly less selective, while $\frac{1}{16}$ are purchased and lead the algorithm to become significantly more selective. In contrast when the budget fraction is $\frac{1}{2}$, this search is much more balanced and has less variance. In comparison, Algorithm \ref{alg:OSLA} does not have these downsides, and outperforms adaptive pacing by $25.9\%$ on average when the budget fraction is $\frac{1}{16}$. Thus Algorithm \ref{alg:OSLA} could have a significant practical impact for bidders in practice, where they typically only have a budget large enough to purchase a small fraction of the ad impressions they might be interested in. 

One may argue that this is not a fair comparison, as the empirical setting provided in this paper does not capture adaptive pacing's ability to react to fundamental marketplace changes. However, in practice it is not clear that this ability is always a net positive. In particular, it is shown in Figure \ref{fig:ecpc} that there are strong seasonal effects throughout the week, where obtaining clicks is cheaper on certain days or during certain times of the day. Because adaptive pacing will smooth the expenditure path over time, this would lead to over/underspending on certain days of the week and also certain hours of the day, when compared to the offline optimum (see Section \ref{sec:dual}). While it has been shown that adaptive pacing can recover a constant competitive ratio in this setting, the competitive ratio is unknown, and the losses could potentially be significant \citep{balseiro2017learning}. In contrast, the algorithm from Example \ref{cor:Gen-RTB} can be trained to avoid these problems.

\section{Conclusions}

In this paper we studied strategies for real-time bidding on Internet advertising exchanges, under the second-price auction format. In particular, we showed a strong connection between selection algorithms for the online knapsack problem with known prices and bidding algorithms for the second-price knapsack problem with unknown prices. This connection has not been leveraged before, and our results yield competitive and readily implementable algorithms for advertising companies. 

To establish these results, we first built an intuition for the connection, by analyzing the offline problem and showing near-optimality of a linear bidding strategy, because of the second-price auction format: in fact the ads recovered are exactly the ones with a value-to-price ratio above some threshold. We then established two specific methods for taking primal and dual online knapsack algorithms, which select ads based on their value and posted price, and turning them into bidding strategies for the second-price knapsack problem, where we do not have a priori knowledge of the other bids (and therefore prices). We give two examples of online knapsack algorithms, adapted to the second-price knapsack setting. In particular, we showed how with an online one-shot learning algorithm, we could use a linear form of bidding to recover a bundle of ads with competitive ratio $1-6\epsilon$, where $\epsilon$ could be small given the context.

Evaluating the adapted algorithm from Example \ref{cor:Gen-RTB} on the iPinYou dataset, we showed that we can recover something close to the optimal bundle of ads as expected, giving empirical support to the findings. In particular, for bidders that are more selective, and only have the budget to purchase $\frac{1}{16}$ of the ads they are interested in, we show that our methods can significantly outperform adaptive pacing, raising the total value of ads purchased by an average of 25.9\%.

While this is not a perfect comparison, as the empirical setting provided in this paper does not capture adaptive pacing's ability to react to fundamental marketplace changes, it was argued in Section \ref{sec:empiricalresults} that this is not always a positive. In particular, there are predictable variations in the value of ads throughout the week, where getting clicks is cheaper during certain days of the week and certain hours of the day. However, adaptive pacing will balance its expenditure rate across all these fluctuations, causing it to either underspend or overspend at certain times. In contrast, the algorithm from Example \ref{cor:Gen-RTB} can be trained to avoid this pitfall.

Therefore the trade-offs in practice, between the work in this paper and adaptive pacing, can be summarized as follows: adaptive pacing requires very little management, but will have some undetermined losses due to predictable seasonal variations, while the online algorithms from Section \ref{sec:onlineselection} can avoid these problems, but must be regularly monitored (and potentially retrained) due to fundamental changes in the marketplace such as changes in competitor behavior. This suggests there might be some instances in practice, where advertisers would prefer the algorithms provided in this paper.

This also suggests there is further room for the development of new algorithms, that combine the best aspects from each of these: being robust to predictable seasonal variations in the attractiveness of ads (over the course of the week), while still being able to adapt to long-term underlying changes in the marketplace.

\section*{Acknowledgement}
The authors would like to thank Haihao Lu. The paper evolved from a Spring 2017 course project at MIT involving him and the authors, that used the iPinYou dataset to evaluate bidding strategies from a literature review of existing work. Haihao also provided thoughtful feedback on the final version of the paper.

\bibliography{biblio.bib}
\bibliographystyle{plainnat}

\newpage
\appendix

\section{Proof of Proposition \ref{thm:DT}}\label{sec:PropositionProof}
In this section we provide a proof of Proposition \ref{thm:DT}. First we introduce some necessary Lemmas, and provide the proof of the Proposition at the end of the section.

Consider (\ref{eq:LP}). As we are trying to minimize a convex piecewise linear function, the optimal value $\lambda^*$ is interpretable as the value-to-price ratio threshold $\frac{v}{b} \geq \lambda$, above which we will select an ad. We provide the following known optimality lemma, similar to \cite{fisher1981lagrangian}.

\begin{lemma} \label{lm: CSC}
Consider the offline knapsack problem (\ref{eq: kpi-opt-trad}) and its linear programming relaxation (\ref{eq:LP}). Let $v_i$ denote the value and $b_i$ the paying price of advertisement $\{i \in I\}$. There exists some constant $\lambda^*$ such that we can recover an optimal solution to formulation (\ref{eq:LP}) using a fractional selection policy $x^*$ such that 
\[ x^*_i = \begin{cases}
1 \text{ if } \frac{v_i}{b_i} > \lambda^*\\
0 \text{ if } \frac{v_i}{b_i} < \lambda^*
\end{cases} \] 
and $x_i^* \in [0,1]$ for the ads with $\frac{v_i}{b_i} = \lambda^*$.
\end{lemma}

\proof{}
The proof relies on complementary slackness. Let $x^*$ denote the optimal solution to the (\ref{eq:LP}), and let $\lambda^*$ and $z^*$ the optimal solution to its dual formulation (\ref{eq:dual}) (these exist by clearly evident strong duality). The following complementary slackness condition entails the described selection policy: \[ \forall i: \: x^*_i (z^*_i - v_i + \lambda^* b_i) =0 \andd (x^*_i - 1) z^*_i = 0 \]
\begin{enumerate}
	\item If $\frac{v_i}{b_i} > \lambda^*$. Since $v_i - \lambda^* b_i > 0$, it follows that $z_i^* > 0$ from the feasibility conditions for ({4}). From the second complementary slackness condition we have $x^*_i = 1$ so we fully purchase the ad.	
	\item If $\frac{v_i}{b_i} < \lambda^*$. From the structure of ({4}) we see that we minimize each choice of $z^*-i$ subject to $z^*_i \geq v_i - \lambda^* b_i)$ and $z^*_i \geq 0$. If $v_i - \lambda^* b_i < 0$ then it follows that we have $z^*_i = 0$ and $z^*_i - v_i + \lambda^* b_i < 0$. From the first complementary slackness condition it follows that $x^*_i = 0$ i.e. we do not purchase the ad.	
	\item Otherwise $ \frac{v_i}{b_i} = \lambda^*$, we purchase some fraction, spending down the remaining budget.
\end{enumerate} 
\endproof

To prove near-optimality of selecting all ads strictly above this value-to-price threshold, we use Assumption \ref{ass:unique}.

Let us consider the formulations (\ref{eq: kpi-opt-trad}) and (\ref{eq:LP}), let us denote the offline dual threshold as $\lambda^*$ and the optimal solution to (\ref{eq:LP}) as $x^*$. From Lemma \ref{lm: CSC}, for the ads where $\frac{v_i}{b_i} > \lambda^*$ we have $x^*_i = 1$, and for the ads where $\frac{v_i}{b_i} < \lambda^*$ we have $x^*_i = 0$. Finally by the uniqueness assumption, we have a single ad indexed by $j$ with $x_j=0$ and $x_i=0$, $\forall i>j$. Define the functions $Z_{LP}(B)$ and $ Z_{IP} (B)$ as the optimal objective values for (\ref{eq:LP}) and (\ref{eq: kpi-opt-trad}) with budget \textit{B}. We derive the following lemma which guarantees the near-optimality of our strategy:
\begin{lemma}\label{lm: IP bound}
	Let $x^*$ be the optimal solution of (\ref{eq:LP}) with budget $B$, described in Lemma \ref{lm: CSC}, where $j$ refers to the unique ad we purchase with $x_j>0$ and $x_i = 0$, $\forall i \text{ s.t. } \frac{v_i}{b_i} < \frac{v_j}{b_j}$. We get the following bound on the optimal selection strategy with budget B:
	\begin{align}
	Z_{IP}(B) - Z_{IP}(B-b_{j}x^*_{j}) \leq  v_{j}
	\end{align}
\end{lemma}
\proof{}
We notice that when our budget is $B+b_{j}(1-x^*_{j})$ rather than $B$ for (\ref{eq:LP}), then the optimal solution from Lemma \ref{lm: CSC} is an integer solution, and thus also optimal for the corresponding integer program. We find a similar result for a budget of $B-b_{j}x^*_{j}$ as well. Combining these results we have:
	\begin{align}
	Z_{LP}(B+b_{j}(1-x^*_{j})) &= Z_{IP}(B+b_{j}(1-x^*_{j})) \nonumber \\&= Z_{IP}(B-b_{j}x^*_{j}) + v_{j} 
	\end{align}
	Then, since feasibility of a solution $x$ stays ensured when provided with a larger budget, we get the inequalities:
	\begin{align}
	Z_{IP}(B) \leq Z_{IP}(B+b_{j}(1-x^*_{j})) = Z_{IP}(B-b_{j}x^*_{j}) + v_{j}
	\end{align}
	We then simplify this inequality to the Lemma's result.
\endproof

When we combine this with Assumption \ref{ass:low}, which states that the value of each ad is small relative to the total value of the optimal bundle (which holds for the context of RTB), we are near-optimal.

So far these results, conditioned on us knowing the paying-price of an ad, are known results for the offline knapsack algorithm. Notably, our selection criteria involves knowledge of the paying price $b_i$, which we can not use in a bidding strategy for the second-price knapsack problem (as $b_i$ will not be known a priori). The following lemma bridges that gap. This lemma is a known result, that has been used in several other papers \citep[e.g.,][]{zhou2008algorithm}.

\begin{lemma}[Scaled linear bid]\label{lm:linear bid}
For any set of ads $\{i \in I\}$ with values $v_i$ and paying prices $b_i$, and for any constant $\lambda^*$, let $\pi^*$ be the linear bidding strategy in the online setting which bids for every impression the amount 
\[ \pi^*_i = \frac{v_i}{\lambda^*} \] Then $\pi^*$ exactly recovers the ads where $\frac{v_i}{b_i} > \lambda^*$ under the second-price auction mechanism.
\end{lemma}
\proof{}
Let us consider the linear form of bidding with parameter $\frac{1}{\lambda^*}$, where we bid the amount $\frac{v_i}{\lambda^*}$ for ad $i$.  The auction is won if and only if the paying price $b_i$ is lower than the bid, i.e. $b_i < \frac{v_i}{\lambda^* } \iff \frac{v_i}{b_i} > \lambda^*$. \\ Therefore because of the second price auction format we find that a linear form of bidding wins the exact set of ads as the dual threshold strategy, and pays the same price.
\endproof

These Lemmas provide the pieces for the formal proof of Proposition \ref{thm:DT}.

\proof{Proof of Proposition \ref{thm:DT}}
Consider the (\ref{eq: kpi-opt-trad}) and (\ref{eq:LP}) formulations from section \ref{sec:dual} and the subsequent notations. With lemma \ref{lm: IP bound}, we know that the total value we can obtain from $Z_{IP}(B)$ is bounded by the value of the linear relaxation with corrected budget and value, i.e. by $Z_{LP} (B-x^*_{i^*}b_{i^*}) + v_{i^*}$.\\
Using lemma \ref{lm:linear bid}, we could recover the same set of ads as in the linear relaxation by using a scaled linear bid with parameter $\frac{1}{\lambda^*}$ for the online problem (\ref{eq: kpi-opt}). Therefore by using this strategy, we are within $\max_i(v_i)$ of the optimal bidding strategy with budget $B$, i.e. the value $Z_{IP}(B)$. This concludes the proof..bib
\endproof

\end{document}